\documentclass[]{article} 

\usepackage{caption}
\usepackage[list=true,labelformat=simple]{subcaption}
\usepackage{subcaption}
\usepackage[square,numbers,sort&compress,comma]{natbib}
\usepackage{siunitx}
\usepackage{amsmath}
\usepackage{amssymb}
\usepackage{caption}
\usepackage{graphicx}
\usepackage{latexsym}
\usepackage{times}
\usepackage[pagewise]{lineno}
\usepackage[export]{adjustbox}

\topmargin - 12pt 
\renewenvironment{abstract}%
              {
               \small
               {\bfseries \abstractname}
               \par
               \vspace{10pt}
              }

\renewcommand\abstractname{Abstract}

\newcommand{\nomenclature}
              [1]
              {
               \bgroup
               \flushleft
               \small\bf
               #1
               \par
               \egroup
              }

\renewcommand{\section}
              [1]
              {
               \bgroup
               \flushleft
               \small\bf
               \refstepcounter{section}
               \arabic{section}. #1
               \par
               \egroup
              }

\renewcommand{\subsection}
              [1]
              {
               \bgroup
               \flushleft
               \small\em
               \refstepcounter{subsection}
               \arabic{section}.
               \arabic{subsection}. #1
               \par
               \egroup
              }

\renewcommand{\subsubsection}
              [1]
              {
               \bgroup
               \flushleft
               \small\em
               \refstepcounter{subsubsection}
               \arabic{section}.
               \arabic{subsection}.
               \arabic{subsubsection}. #1
               \par
               \egroup
              }

  \newcommand{\acknowledgement}
              [1]
              {
               \bgroup
               \flushleft
               \small\bf
               #1
               \par
               \egroup
              }

  \newcommand{\sectionbib}
              [1]
              {
               \bgroup
               \flushleft
               \small\bf
               #1
               \par
               \egroup
              }

\begin{document}

\title{\LARGE The critical conditions for the re-ignition and detonation formation from Mach reflections of curved decaying shocks}

\author{{\large Farzane Zangene$^{a,*}$, Matei Radulescu$^{a}$}\\[10pt]
        {\footnotesize \em $^a$Department of Mechanical Engineering, University of Ottawa
        , Ottawa, ON K1N6N5, Canada}\\[-5pt]}

\date{}


\small
\baselineskip 10pt


\vspace{50pt}
\maketitle
\vspace{40pt}
\rule{\textwidth}{0.5pt}
\begin{abstract} 
The objective of this study is to determine the critical conditions for a detonation wave formation following a Mach reflection of two incident shocks. This problem is central to the propagation mechanism of cellular detonations, where such periodic reflections occur continuously. A new experimental technique is introduced that permits the isolation of this Mach reflection. The technique is inspired by that of White (1963) and uses a detonation passing through a bifurcated converging-diverging nozzle to reproducibly generate the desired shock reflection at its exit, where the two transmitted mis-aligned incident shocks interact. The collision process was monitored with high-speed Schlieren videos permitting to measure the strength, decay rate and curvatures of the incident and Mach waves. The experiments were performed in mixtures of $\text{C}\text{H}_\text{4}/\text{2}\text{O}_\text{2}$ and $\text{2}\text{H}_\text{2}/\text{O}_\text{2}/\text{2}\text{Ar}$, which span the degree of cellular regularity of detonations in reactive gases. Three distinct regimes of transmission were identified: detonation formation, decaying Mach shock followed by a flame, and inert Mach shock with no re-ignition. For both mixtures, the condition separating the re-ignition and no re-ignition was very well predicted by the critical decay rate ignition theory of Eckett et al. On the other hand, the transition between detonation formation and ignition was found to correlate much better with the critical curvature concept of quasi-steady curved detonations of Kasimov and Stewart. Both experiments in the regular and irregular mixtures were found in excellent agreement with this criterion, established on the basis of experimental velocity-curvature data available in the literature for these mixtures. For the hydrogen detonations, this criterion was also in agreement with the critical curvature predicted by ZND theory. For the methane detonations, order of magnitude disagreement was observed with the ZND model prediction, further confirming the propensity of irregular cellular detonations to propagate under higher curvatures than predicted by laminar detonation theory. 
\end{abstract}
\vspace{10pt}
\parbox{1.0\textwidth}{\footnotesize {\em Keywords:} Detonation formation; Shocks reflection; Decay rate model; Weakly non-steady curved detonations.}
\rule{\textwidth}{0.5pt}
\vspace{10pt}


\clearpage


\section{Introduction\label{sec:introduction}} \addvspace{10pt}
It is well known that all self-sustaining detonations in gases have a three-dimensional cellular structure \cite{Lee:1984}. These interactions are believed to control the propagation mechanism of detonations, as they periodically generate highly overdriven detonations well above the Chapman-Jouguet (CJ) speed, which then decay through the cycle and reform from new triple point collisions \cite{Fickett:1979}.  In the propagation of detonation waves, especially in the vicinity of marginal propagation conditions, two distinct behaviors are observed within the cell. In one scenario, following two triple points collision, a local detonation wave is generated, in which the Mach shock is coupled with the reaction wave behind it, displaying a fine cellular structure. Conversely, in the other scenario, there is minimal exothermicity behind the newly formed Mach shock, and the majority of the gas accumulates as unreacted, subsequently undergoing combustion within a turbulent flame \cite{radulescu2003propagation, Radulescu:2007}. Currently, it remains unclear what conditions lead to the local reformation of detonation from triple shock reflections and what triggers the occurrence of either of these two behaviors. The ability to predict when each of these scenarios occurs is crucial for understanding critical phenomena related to detonation initiation and failure, such as direct initiation, detonation diffraction, etc.

Previous attempts at identifying a canonical configuration to study this phenomenon have yielded limited success.  In situ study of Mach reflections in detonations is difficult owing to the small scale and stochasticity of the phenomenon.  Attempts to generate reproducible Mach reflections in the wakes of obstacles placed in the path of detonations have also yielded limited success, due to the presence of both regular and Mach reflections in the same experiment \cite{bhattacharjee2013detonation}.  In the present study, we formulate a novel configuration to isolate this Mach reflection.  The technique is inspired by that of White \cite{white1963structure} and uses a detonation passing through a bifurcated converging-diverging nozzle to reproducibly generate the desired shock reflection at its exit, where the two transmitted mis-aligned incident shocks interact. We report experiments using this technique.  We observe three distinct regimes of transmission: Mach reflection without re-ignition, Mach reflection with re-ignition and Mach reflection with the generation of coupled detonation waves generating a fine-scale transverse wave structure.   

The conditions separating the various regimes of transmission are discussed in terms of two conceptual models: the critical expansion model for gas ignition \cite{lundstrom1969influence, eckett2000role, Austin:2003,  radulescu2010critical, cheevers2022ignition} and the critical sustenance criteria of weakly curved and weakly non-steady detonations anticipated in Detonation Shock Dynamics theory \cite{yao1996dynamics}.  

\section{Experiments\label{sec:unnum}} \addvspace{10pt}
\subsection{Experimental set-up\label{subsec:subsection}} 
\begin{figure*}
\centering
\includegraphics[scale=0.95]{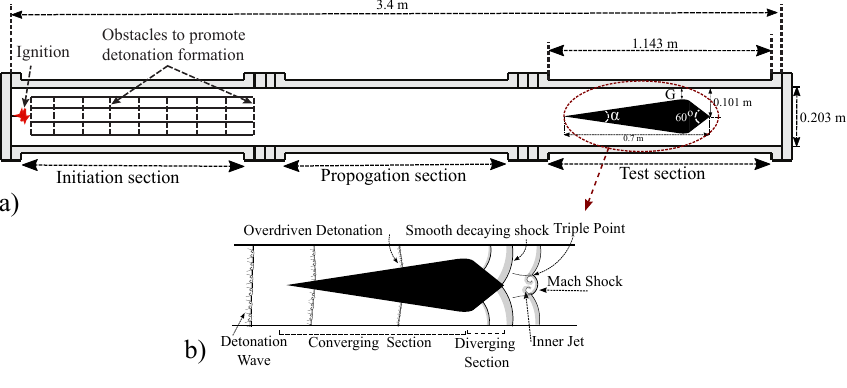}
\caption{The schematic of the experimental apparatus to study the Mach shock formation in a reactive gas from shock reflections. The angle $\alpha$ is set to either 15$^\circ$ or 20$^\circ$, corresponding to $G$ values of 0.0265-m and 0.0065-m, respectively.}
\label{fig:ST}
\end{figure*}

\addvspace{10pt}
The experiments were performed in a 3.4-m-long shock tube, with 0.019-m-thickness and 0.203-m-height \cite{Bhattacharjee}. The schematic illustrating the experimental set-up is shown in Fig.\ \ref{fig:ST}a. A 0.7-m diamond shape object made of aluminum was inserted into the test section to create an inverted converging-diverging nozzle with the top and bottom walls of the shock tube, symmetrically.  

In Fig.\ \ref{fig:ST}a, the variable $G$ denotes the channel height before the detonation diffraction occurs. Two different objects were specifically designed to control the structure of the two decaying shocks through the diffraction process, creating gap sizes of 0.0265 m and 0.0065 m. The angle of the converging section, denoted by $\alpha$, was chosen to be 15$^{\circ}$ for the gap size of 0.0265 m and 20$^{\circ}$ for the gap size of 0.0065 m. This selection aimed to provide two smooth overdriven detonation waves without any kink on the structure as documented in a separate study \cite{akbar1997mach}.  To ensure a smooth transition, the object's throat connecting the converging and diverging channels had a rounded tip with a radius of curvature equal to the gap size.

In the diverging section, the detonation is made to recover an idealized quasi-laminar structure of a shock followed by a reaction wave. As sketched in Fig.\ \ref{fig:ST}b, the two symmetry decaying shocks from the top and bottom channels interact at the apex of the diamond and generate a Mach reflection. The angle of the diverging section was maintained 30$^{\circ}$.  This selection was motivated by the observed tendency of detonation Mach reflections to have a 30$^{\circ}$ angle of incidence of the incident shocks \cite{Austin:2003, ZangeneTP:2023} 

The propagation and reflection process was visualized using the Schlieren technique using 30 cm diameter field mirrors, a high-speed Phantom v1210 camera and a continuous white light source from an incandescent automotive light bulb. The inter-frame times were 13.6 $\mu s$ and 12.9  $\mu s$  for resolutions of 384 $*$ 304 $\text{px}^2$ and 384 $*$ 288 $\text{px}^2$, respectively. 

The two mixtures studied were a stoichiometric mixture of methane and oxygen ($\text{C}\text{H}_\text{4}/\text{2}\text{O}_\text{2}$) and a stoichiometric mixture of hydrogen-oxygen diluted with argon ($\text{2}\text{H}_\text{2}/\text{O}_\text{2}/\text{2}\text{Ar}$). For each mixture, its sensitivity was controlled by changing the initial pressure of the test gas.    These two mixtures were selected by their substantial differences in cellular regularity. Their characteristic properties are also summarized in Table \ref{tab:mixture}. Their adiabatic index evaluated at the Von-Neumann state, $\gamma_{VN}$, and the detonation stability parameter  \cite{radulescu2003propagation}, $\chi$, change over a sufficiently wide range. All the reported experiments are conducted at an initial temperature of 293 \si{\kelvin}.

\begin{table}
\centering
\caption{The experimental test gases and their properties.}

\begin{tabular}{lcccc}
\hline
 Mixture &$P_0$ [\si{\kilo\pascal}]& $\gamma_{VN}$ 	&	$\chi$ \\
\hline
 $\text{C}\text{H}_\text{4}/\text{2}\text{O}_\text{2}$&  15 &	1.17&	510  \\
 $\text{2}\text{H}_\text{2}/\text{O}_\text{2}/\text{2}\text{Ar}$&15	&1.4	&3.7 	\\
\hline
\end{tabular}
\label{tab:mixture}
\end{table}
 
\subsection{Experimental results: the role of the CD nozzle\label{subsec:subsection}} \addvspace{10pt}

\begin{figure*}
\centering
\begin{subfigure}{0.45\columnwidth}
\centering
    \includegraphics[width=\textwidth]{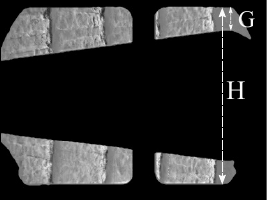}
    \caption{$\text{CH}_4/\text{2O}_2$ at $P_0$= 17 \si{\kilo\pascal}.}
\end{subfigure}
\hfill
\begin{subfigure}{0.45\columnwidth}
\centering
    \includegraphics[width=\textwidth]{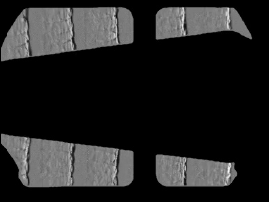}
    \caption{$\text{2H}_2/\text{O}_2/\text{2Ar}$ at $P_0$= 9.6 \si{\kilo\pascal}. } 
\end{subfigure} 
\caption{Composite Schlieren images of the formation of the overdriven detonation through the converging channels. The height of the channel, H, is 0.203-m and the gap size, G, is 0.0265-m.}
\label{fig:CH4-H2back}
\end{figure*}

The role of the converging section was to overdrive the detonation.  Figure \ref{fig:CH4-H2back} depicts the detonation structures of the two mixtures during this compression.  The key feature to note is the absence of a Mach type reflection at these small angles of incidence.  This appears to be a signature of reactive Mach reflections, as recently studied by Short et al.\ \cite{short2023icders}. This desirable feature makes the technique adequate, as it does not introduce strong reflections that will persist in the diverging section of the nozzle. 

Nevertheless, the cellular structures persist on the front of the overdriven detonations.  The methane–oxygen detonation retains the spotty structure indicative of non-reacted pockets, as well as its long-lived transverse waves.  These features are less pronounced in the hydrogen–oxygen–argon mixture. In these experiments, the detonation speed increases to approximately 1.1$D_{CJ}$ and 1.06$D_{CJ}$ at the throat in the methane-oxygen and hydrogen-oxygen-argon mixtures, respectively.

The success of our technique requires that the detonation transmission in the enlarging section is subcritical, giving rise to a decaying shock followed by a decoupled reaction zone.  The criterion for this was found in good agreement with detonation diffraction phenomenology.  By applying equation (12) from \cite{Xiao:2021}, which establishes a relationship between the exponential sensitivity of cell size and induction zone length to velocity deficit in the strong shock limit, we estimated the cell size of the overdriven detonation, denoted as $\lambda_\text{Overdriven}$, at the end of the converging section. Our findings indicate that the conditions for the wave to undergo attenuation after the transition to the diverging section align with the critical channel height in diffraction problems. The reported critical channel height ranges from $G/\lambda\approx3-10$, with variations dependent on mixture irregularity and the dimensions of the opening.
\subsection{Experimental results: the Mach reflection\label{subsec:subsection}} \addvspace{10pt}
In our experimental observations, we identified three distinct regimes immediately following the reflection of the two incident shocks. These regimes are the formation of a detonation wave, the creation of a decaying Mach shock accompanied by a reaction front, and the generation of an inert Mach shock with no ignition occurring behind it.

\begin{figure*}
\centering
\begin{subfigure}{\columnwidth}
    \includegraphics[width=\textwidth]{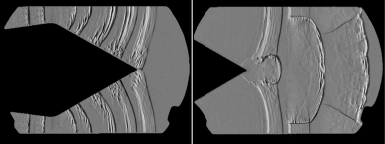}
    \caption{Detonation formation, $P_0$= 9.6 \si{\kilo\pascal}}
    \label{fig:H2-Det-Exp}
\end{subfigure}
\hfill
\begin{subfigure}{\columnwidth}
    \includegraphics[width=\textwidth]{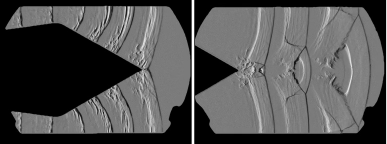}
    \caption{Decaying Mach shock followed by a flame, $P_0$= 9 \si{\kilo\pascal}. } 
\label{fig:H2-Ig-Exp}
\end{subfigure} 
\hfill
\begin{subfigure}{\columnwidth}
    \includegraphics[width=\textwidth]{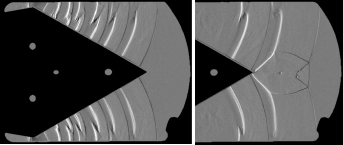}
    \caption{Inert Mach shock with no re-ignition, $P_0$= 9.6 \si{\kilo\pascal}. } 
\label{fig:H2-NoIg-Exp}
\end{subfigure} 
\caption{Composite Schlieren images of the formation of the three distinctive regimes in the $\text{2H}_2/\text{O}_2/\text{2Ar}$ mixture.}
\label{fig:}
\end{figure*}

Figure \ref{fig:H2-Det-Exp} illustrates the diffraction of the detonation in the diverging channel followed by the formation of Mach shock from shock reflection. Multiple frames are overlaid in the same image to illustrate the evolution of the two incident shocks over time before the reflection. Schlieren photographs, capturing density gradients, distinctly reveal the boundary between burned and shocked yet non-burned gases. The reaction zone consistently lags behind the lead shock. Following the interaction of the two decaying shocks on the plane of symmetry, a strong overdriven detonation is formed and continues propagating along the channel. The cellular structure in the newly formed detonation is clearly visible. 

In a slightly less sensitive mixture, detonation fails to form from the reflection, signifying the close proximity to the transition between the two regimes.  An example is shown in Fig.\ \ref{fig:H2-Ig-Exp}.  The transmitted Mach reflection takes the form of a decaying shock followed by a reaction zone trailing far behind.  The absence of combustion coupled to the shock front leads to the accumulation of unburned gas behind the decaying shock over time. 
\begin{figure*}
\centering
\includegraphics[scale=0.6]{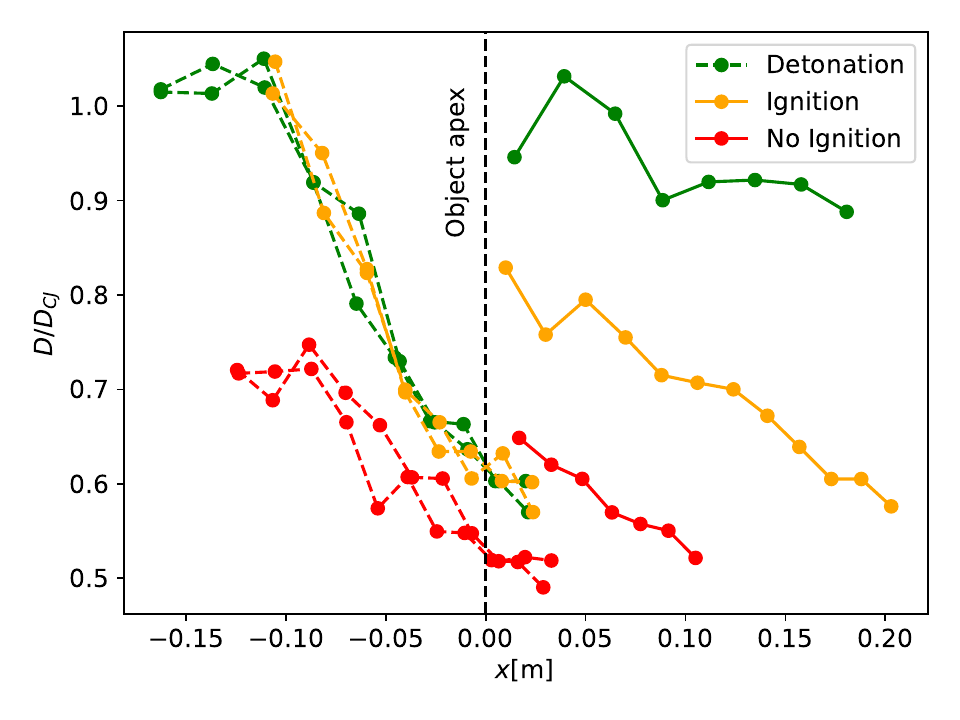}
\caption{The evolution of the velocity deficit with distance for the two decaying incident shocks on the top and bottom walls of the shock tube (represented by dashed lines) and the Mach shock on the centerline (illustrated by solid lines) in the $\text{2H}_2/\text{O}_2/\text{2Ar}$ mixture.}
\label{fig:DXcenterH2}
\end{figure*}
Altering the geometry to reduce the gap size by a factor of four ($G$ = 0.0065 m) significantly accelerates the decoupling between the shock and reaction front in the diverging section, as evident in Fig.\ref{fig:H2-NoIg-Exp}. The two incident shocks decay much more rapidly, observable in the Schlieren images as the unreacted region behind the shock front progressively increases. At the apex of the object where the two incident shocks interact, an inert Mach shock is formed, indicating the absence of ignition behind the shock front. The accumulation of unreacted gas behind the shock serves as a signature of the shock's weakness compared to the other scenarios.

The strength of the decaying shocks prior to the collision and the strength of the Mach shocks for the three previous examples are shown in Fig.\ \ref{fig:DXcenterH2}.  The speed of the incident shock waves decaying in the diverging section was measured along the top and bottom walls.  The speed of the Mach shock was measured along the centerline.  It is evident that the local speed profile undergoes significant deceleration along the diverging section, decreasing the velocity from 1.15$D_{CJ}$ to 0.6$D_{CJ}$ for both the cases of detonation and decoupled Mach shock-reaction front. This deceleration is even more pronounced in the no ignition case, particularly when a small gap size is involved, dropping the velocity to 0.5$D_{CJ}$. Following the interaction of the two decaying shocks, the velocity of the newly formed Mach shock is approximately 0.92$D_{CJ}$ for the detonation formation and due to the effect of combustion it rapidly increases to 1.15$D_{CJ}$ and continues to propagate at around 0.9$D_{CJ}$. Conversely, in scenarios where detonation did not occur, the Mach shock velocity is 0.82$D_{CJ}$ and continues to decay to lower velocities as it is decoupled from the reaction front. In the case of no ignition, where the two incident shocks exhibit relatively low strength, the Mach shock's velocity only experiences a 10$\%$ increase after reflection, reaching 0.65$D_{CJ}$. Subsequently, it continues to decay rapidly, given its nature as a weak inert shock.

\begin{figure*}
\centering
\begin{subfigure}{\columnwidth}
    \includegraphics[width=\textwidth]{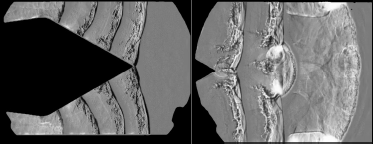}
    \caption{Detonation formation, $P_0$= 19 \si{\kilo\pascal}}
\label{fig:CH4-Det}
\end{subfigure}
\hfill
\begin{subfigure}{\columnwidth}
    \includegraphics[width=\textwidth]{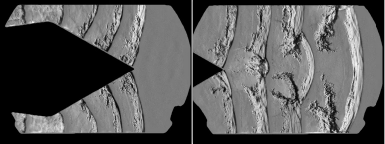}
    \caption{Decaying Mach shock followed by a flame, $P_0$= 19 \si{\kilo\pascal} } 
\label{fig:CH4-Ig}
\end{subfigure} 
\hfill
\begin{subfigure}{\columnwidth}
    \includegraphics[width=\textwidth]{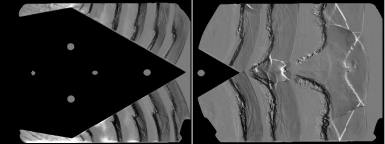}
    \caption{Inert Mach shock with no re-ignition, $P_0$= 17.2 \si{\kilo\pascal} } 
\label{fig:CH4-NoIg}
\end{subfigure} 
\caption{Composite Schlieren images of the formation of the three distinctive regimes in the $\text{CH}_4/\text{2O}_2$ mixture.}
\end{figure*}

The experiments conducted in the methane-oxygen mixtures showed qualitatively similar behaviour.  Fig.\ \ref{fig:CH4-Det} illustrates a reflection giving rise to a detonation.  Fig.\ \ref{fig:CH4-Ig} illustrates a reflection giving rise to ignition and decay, whereas Fig.\ \ref{fig:CH4-NoIg} shows a case with quenched ignition behind the decaying shock. Both the experiments in Figs. \ref{fig:CH4-Det} and \ref{fig:CH4-Ig} were obtained at the same operating conditions, hence reflect the critical conditions separating the two regimes.   Due to the cellular irregularity and highly sensitive chemistry of the methane-oxygen mixture, the reaction front is characterized by significant hydrodynamic fluctuations and unreacted pockets, which remain even in the enlarging sections of the channel. This stands in contrast to the mostly laminar reaction front observed in the hydrogen-oxygen-argon mixture as it decays in the diverging section. 

The detonation initiation case of Fig.\ \ref{fig:CH4-Det} displays the characteristic very fine cellular structure on the front and transverse detonations.  These can be identified by the very high levels of light emission overwhelming the Schlieren signal.  A much finer-scale inner structure than the hydrogen detonations is clearly evident.  
In contrast, Fig.\ \ref{fig:CH4-Ig} shows a decaying Mach shock followed by a mostly laminar reaction front. The transverse waves are non-reactive in this case, and long tongues of unburned gas are apparent, accompanied by a very fine scale structure.  At the nascence of this Mach reflection, ignition and forward jet entrainment are obvious - see the high-speed videos available as supplementary material.  These take the same form as the jet structures observed in our previous studies using round obstacles to induce the reflection process \cite{bhattacharjee2013detonation}.
\begin{figure*}
\centering
\includegraphics[scale=0.6]{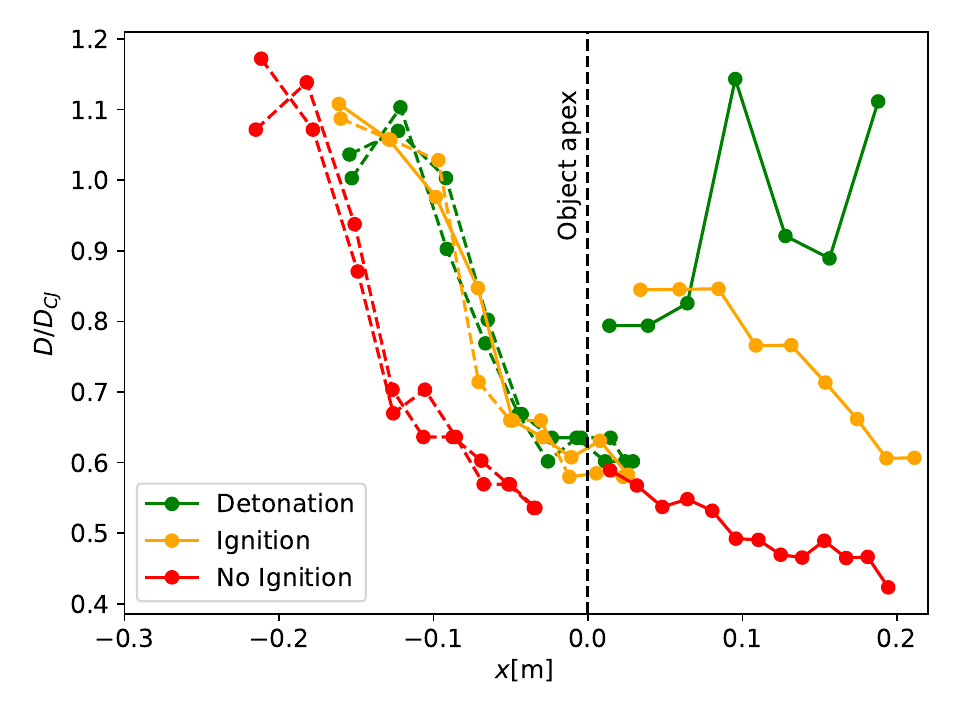}
\caption{The evolution of the velocity deficit with distance for the two decaying incident shocks on the top and bottom walls of the shock tube (represented by dashed lines) and the Mach shock on the centerline (illustrated by solid lines) in the $\text{CH}_4/\text{2O}_2$ mixture.}
\label{fig:DXcenterCH4}
\end{figure*}

For the experiment illustrated in Fig.\ \ref{fig:CH4-NoIg} for a reduced gap size, there is a pronounced quenching effect on the two incoming incident shocks as they diffract. These decaying shocks are very smooth, with a combustion zone decoupled in the back.  The resulting Mach reflection did not re-ignite the gas.  Instead, the detailed structure of a double Mach reflection followed by a forward-facing jet is clearly observable.  Shear layer instabilities along the vortex line joining the jet head are also clearly discernible.  

Figure \ref{fig:DXcenterCH4} shows the velocity evolution along the shock tube walls in the diverging channel and the center line after the reflection for the methane-oxygen mixture. In all three scenarios of detonation, ignition, and no ignition, the velocity transitions from overdriven detonation to a weak shock before the collisions of the two triple points. However, in the first two cases, after the collision, it increases to 0.8 $D_{CJ}$, while in the no ignition case, the reflection of two extremely weak shocks forms a weaker Mach shock of approximately 0.6 $D_{CJ}$. The disparities in velocities of the Mach shocks between the ignition and detonation scenarios are not as prominent as observed in the hydrogen-oxygen-argon cases. This may be attributed to the presence of the forward jet behind the Mach shock, turbulent reaction front, and other hydrodynamic instabilities in the more unstable mixture of methane-oxygen.  The sole differentiation between the two regimes based on speed alone makes the difference between the locally gasdynamically coupled detonation and the shock followed by turbulent combustion less obvious.  Nevertheless, the two regimes are markedly different in the structure of the front.  The detonation initiation is accompanied by a very fine scale cellular structure, while the shock-turbulent flame is not.
\section{Prediction of detonation, ignition and no ignition scenarios\label{sec:unnum}} \addvspace{10pt}
In the following two sections, the conditions that differentiate between the various transmission regimes are explored using two conceptual models related to detonation dynamics: the critical ignition due to volumetric expansion behind decaying shocks and critical conditions for establishing local gasdynamic choking and quasi-steady propagation.
\subsection{Model for critical ignition behind decaying Mach shock\label{subsec:subsection}} \addvspace{10pt}
The quenching of chemical reactions in the presence of strong gas dynamic expansions has been identified by numerous researchers as an important mechanism controlling ignition behind decaying shocks \cite{lundstrom1969influence, eckett2000role, vidal1999analysis, radulescu2010critical}. A summary of the state of the art is provided by Cheevers and Radulescu \cite{cheevers2022ignition}.   The main contribution is that by Eckett et al.\ \cite{eckett2000role}, who analyzed the ignition process along a particle path and identified a critical expansion rate, which prevents ignition.  Extinction is then predicted to occur when 
\begin{equation}
\zeta=\frac{Ea}{RT_{s}}(\gamma-1)\frac{t_{i}}{t_{exp}}>1
\label{eq:ig}
\end{equation}
where $\frac{Ea}{RT_{s}}$ is the dimensionless activation energy, $T_s$ is the post-shock temperature, $t_{i}$ is nominal ignition delay time in the absence of expansion and ${t_{exp}}$ is the characteristic time of the expansion along a particle path. 

To link this criterion to ignition behind a decaying shock, it is sufficient to determine the relation between the expansion rate along a particle path and the shock speed, decay rate and curvature. Radulescu \cite{radulescu2010critical} established this relation by using the shock change equations.  For a strong shock, the relation is:
\begin{equation}
t_{exp}^{-1}=\frac{1}{\rho}\frac{D\rho}{Dt}=\frac{6\dot{D}}{(\gamma+1) D}+\frac{2D\kappa (\gamma-1)}{(\gamma+1)^2}
\label{eq:te}
\end{equation}
This relation generalizes the result of Eckett et al. obtained for planar shocks, where only the first term in \eqref{eq:te} appeared in their analysis.  

To test this ignition model in our experiments, we require the shock speed of the Mach shock $D$, its decay rate $\dot{D}$, its curvature $\kappa$ and the ignition delay behind the inert Mach shock $t_{i}$.  We determine the strength of the Mach shock in our experiments by employing real gas calculations of standard Mach reflections, assuming a flat Mach shock. We provide the details of the computational procedure elsewhere \cite{ZangeneTP:2023}.  The calculation requires the speed of the incident shock, which we extract from our experiments as the average velocity of the two incoming incident shocks, and the reflection angle of 30$^{\circ}$ of our experiments.  With the strength of the Mach shock estimated, the nominal ignition delay time was determined using constant-volume homogeneous reactor calculations from Cantera \cite{cantera}.

To determine the deceleration of the Mach shock, we make use of the relation 
\begin{equation}
\frac{\dot{D}}{D}=\frac{\mathrm{d}D}{\mathrm{d}x}
\label{eq:ddot}
\end{equation}
Our monitoring of the Mach reflection in the inert regime, and analysis of our results for inert shock reflections indicate that $\frac{\dot{D}}{D}$ remains approximately constant through the reflection.  The constancy of the expansion rate across the reflection can also be observed in the results of Figs.\ \ref{fig:DXcenterH2} and \ref{fig:DXcenterCH4} for the inert reflections.  
	
Finally, we also measured the curvature of the Mach shocks and found that its contribution to Eq. \eqref{eq:te} was, in fact, negligible for the conditions of the experiment.  Consequently, the influence of curvature was not further included in the model prediction. 

\begin{figure*}
\centering
\includegraphics[scale=0.7]{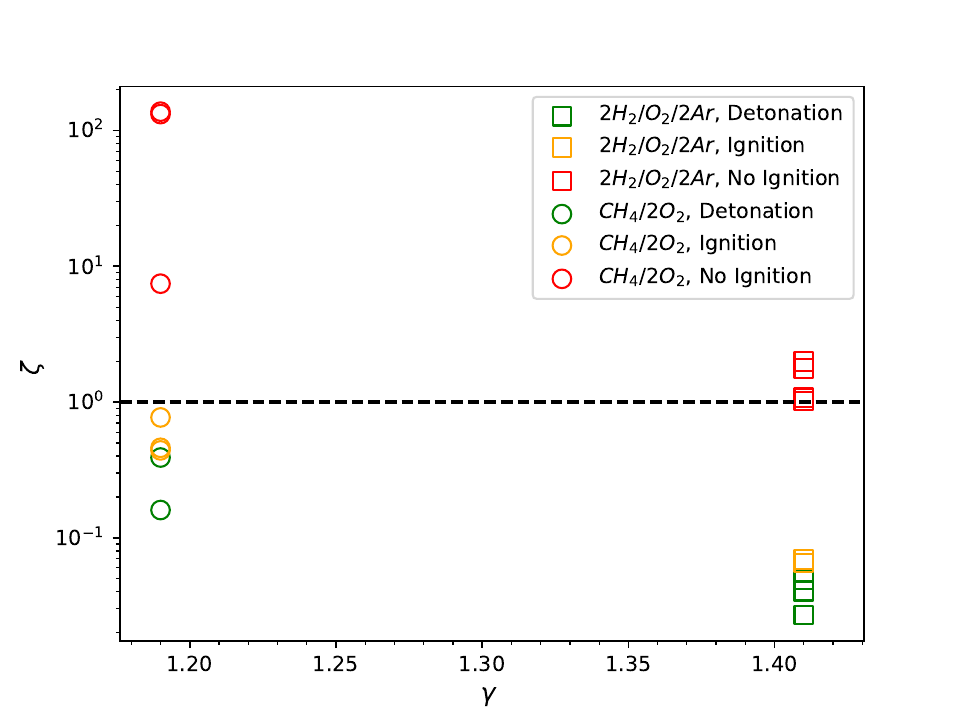}
\caption{The decay rate model prediction, $\zeta$, plotted against the compressibility factor, $\gamma$.  The graph includes the results of eight experiments in the $\text{CH}_4/\text{O}_2$ mixture ($P_0=$7–21 \si{\kilo\pascal}) and ten experiments with the $\text{2H}_2/\text{O}_2/\text{2Ar}$ mixture ($P_0=$8–9.8 \si{\kilo\pascal}).}
\label{fig:Model}
\end{figure*}

Fig.\ \ref{fig:Model} provides a summary of the experimentally inferred decay rate coefficients $\zeta$ of the Mach shock and comparison to Eckett's criterion given by Eq. \eqref{eq:ig}.  As can be verified, the criterion serves as an excellent indicator for the boundary between the experiments where ignition was observed behind the lead shock, and experiments where ignition was not observed.  We can thus conclude that the physical mechanism leading to the quenching of the gas and the absence of ignition is excessive cooling in the induction zone resulting from the volumetric expansion induced by the deceleration of the leading shock. More correctly said, it is due to the volumetric expansion that is also \textit{responsible} for the shock decay - brought about by the diverging geometry controlling the incident shock, which the Mach shock inherits from its predecessor.

We note however that the ignition criterion fails to predict the transition between detonation formation and no detonation (marked by the transition between green and yellow circles).  For the hydrogen mixture, it can be seen that the criterion is more than an order of magnitude different from the experimental results.  The ignition quenching criterion is not adequate.  This discrepancy cannot be attributed to experimental or data reduction error, estimated at a factor of 2, since the hydrogen experiments were sufficiently well resolved to accurately infer the shock speed, its decay rate and ignition behind it using available chemical kinetic models for hydrogen.

\subsection{Model for critical conditions for detonation initiation\label{subsec:subsection}} \addvspace{10pt}
As discussed in the literature, it is believed that self-sustaining detonations require gas-dynamic choking for sustenance and amplification, as conceptualized by Detonation Shock Dynamics theoretical framework for weakly curved, weakly non-steady detonations \cite{bdzil2007, yao1996dynamics, kasimov2005asymptotic, vidal2009critical}. Curvature and non-steadiness in the quasi-steady model account for losses that bring about substantial detonation speed deficits.  The quasi-steady relation between detonation deficit and curvature acts as an attractor for critically forming detonations \cite{kasimov2005asymptotic, vidal2009critical}.  Previous studies of propagating detonations suggested that the early parts of the shock decay inside the cell can be described by this DSD framework \cite{jackson2019intrinsic}. 

\begin{figure*}
\centering
\includegraphics[scale=0.5]{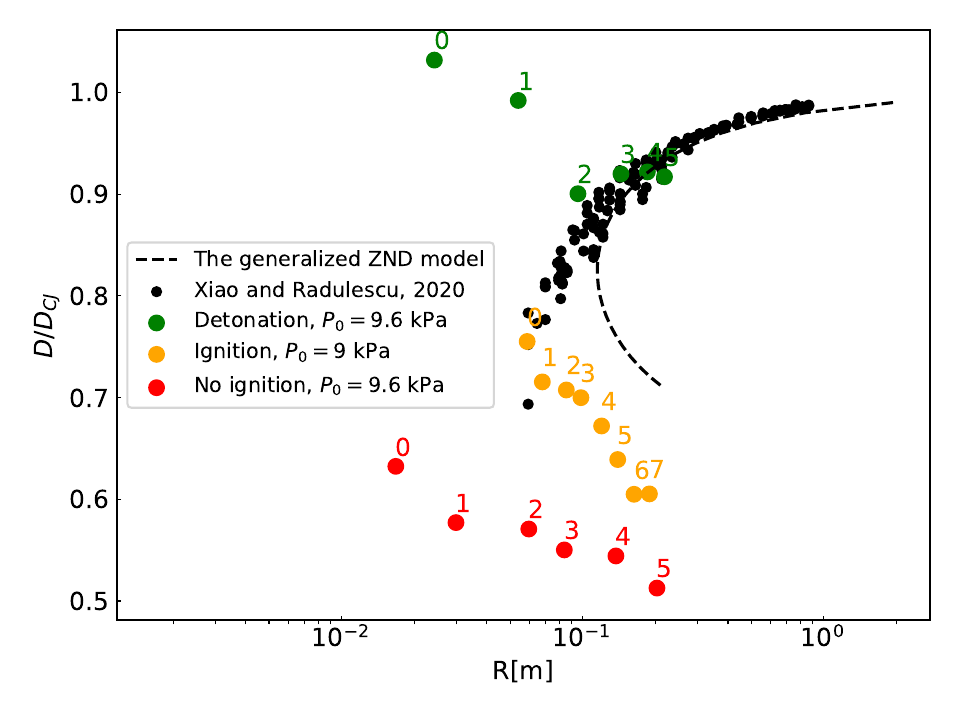}
\caption{The velocity deficit versus radius data obtained from experiments conducted in this study, quasi-steady-state detonation experiments \cite{xiao:2020dynamics}, and predictions derived from the Generalized ZND model. The test gas is $\text{2}\text{H}_\text{2}/\text{O}_\text{2}/\text{2}\text{Ar}$.}
\label{fig:DKH2}
\end{figure*}

\begin{figure*}
\centering
\includegraphics[scale=0.5]{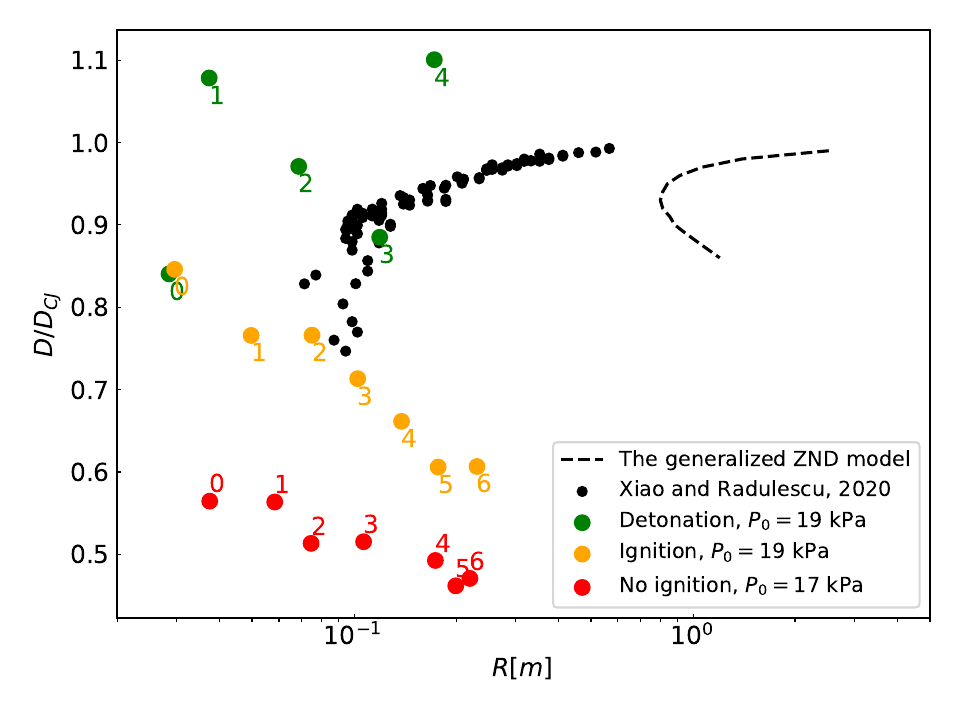}
\caption{The velocity deficit versus radius data obtained from experiments conducted in this study, quasi-steady-state detonation experiments \cite{Xiao:2020}, and predictions derived from the Generalized ZND model. The test gas is $\text{C}\text{H}_\text{4}/\text{2}\text{O}_\text{2}$.}
\label{fig:DKCH4}
\end{figure*}

To explore the usefulness of this concept \cite{kasimov2005asymptotic, vidal2009critical} in predicting the criticality observed in our experiments for detonations formed or not formed, we extracted the detonation speed evolution as a function of its local radius of curvature $R=1/\kappa$.  The curvatures were obtained by curve-fitting the entire shock shape of the Mach shock obtained experimentally.  Figures \ref{fig:DKH2} and \ref{fig:DKCH4} show the data obtained for the same experiments analyzed above for the three regimes of reflection. Each data point represents the average between two consecutive frames, recorded from the first frame after the formation of the Mach shock, at which the extraction of radius data became feasible. 

To make the comparison with the quasi-steady curvature response, we also plot the prediction using ZND theory and the experimentally measured speed-curvature data obtained in our previous studies using the exponential horn technique \cite{xiao:2020dynamics, Xiao:2020}. The results obtained for the hydrogen mixture shown in Fig.\ \ref{fig:DKH2} are striking. The critical experiment leading to detonation formation (green circles) suggests that the reflection gives rise to a near CJ detonation that rapidly decays to its weakly curved evolution given by the ZND model and experimentally determined curvature response.  On the other hand, the decaying Mach shock experiment (yellow circles) indicates that the Mach shock evolves below the bottom branch of the $D(R)$ curve, which implies decoupling dynamics in DSD theory \cite{kasimov2005asymptotic, vidal1999analysis}.  Our results are thus in perfect agreement with DSD theory.   

The results obtained for the methane-oxygen system shown in Fig.\ \ref{fig:DKCH4} indicate similar dynamics, albeit less clearly.  The case of the Mack shock followed by a decoupled reaction zone decays below the critical turning point in the experimentally determined speed-curvature curve (black points).  The dynamics of the detonation forming show significant oscillations while approaching the top portion of the experimentally determined speed-curvature curve.  Nevertheless, the bifurcation between the initiating case and failure case appears to occur for the smaller radius of curvature on the order of a few millimetres.   This approaches the experimental limit at which we can estimate the curvature of the front.  Clearly, experiments with higher resolution and better control of the incident wave structure are required in this case to reach less ambiguous results.  Nevertheless, what we can conclude with certainty is the inability of the curved ZND model to make a useful prediction in this case, since the bifurcation in the experiments occurs at a radius of curvature an order of magnitude smaller than anticipated from laminar theory.  This again stresses the inability of the laminar model to capture the dynamics of the nascent detonations in very irregular mixtures.  The reason for this discrepancy is believed to be the role of the fine-scale structure developing very fast on the surface of these detonations, invalidating the one-dimensional laminar assumption. 

\section{Conclusion\label{sec:unnum}} \addvspace{10pt}
The current study introduced a novel technique to effectively isolate the process of reactive Mach shock formation from the reflection of two incident shocks. Three distinct regimes were successfully identified: detonation formation, ignition occurring behind the decaying Mach shock, and Mach shock with no ignition.  Utilizing high-speed Schlieren videos, measurements were taken to quantify the strength, decay rate, and curvatures of both incident and Mach shocks.

We demonstrated that the decay rate model, incorporating the expansion behind the decaying shock, effectively predicts the critical conditions that differentiate between ignition and no-ignition regimes in both mixtures. Critical conditions that differentiate between the regimes of detonation initiation and failure are compatible with the theory of weakly non-steady curved detonations. The formation of a quasi-steady detonation wave can be explained by the establishment of a sonic surface within the structure of detonation waves as it reaches the critical curvature point. While the critical curvature was well predicted by laminar ZND theory for weakly curved detonations for the regular structure hydrogen-oxygen-argon mixture, this was not the case for the much more irregular detonations in methane-oxygen.  For this mixture, agreement was much better with the mean detonation-curvature relation obtained for cellular detonations. It can be speculated that the fine sub-scale cellular structure promotes the propagation of these detonations.\\
\acknowledgement{Acknowledgments} \addvspace{10pt}
This work was supported by AFOSR grant FA9550-23-1-0214, with Dr.\ Chiping Li as program monitor and the NSERC Discovery Grant "Predictability of detonation wave dynamics in gases: experiment and model development".
\acknowledgement{Supplementary material} \addvspace{10pt}
Supplementary videos associated with this article can be found in the online version.


 \footnotesize
 \baselineskip 9pt


\bibliographystyle{pci}
\bibliography{RefWFD2023}


\newpage

\small
\baselineskip 10pt



\end{document}